\documentclass[a4paper]{article}
\usepackage{xcolor}
\usepackage{INTERSPEECH2021}
\usepackage{graphicx}

\renewcommand\vec{\mathbf}
\newcommand{\ee}{\text{e2e}}
\newcommand{\id}{\text{ID}}
\newcommand{\od}{\text{OOD}}
\newcommand{\nent}{\text{NE}}

\title{Contextual Density Ratio for Language Model Biasing of Sequence to Sequence ASR Systems}
\name{Jes\'{u}s Andr\'{e}s-Ferrer\textsuperscript{1,2}, Dario Albesano\textsuperscript{1,3}, Puming Zhan\textsuperscript{4},  Paul Vozila\textsuperscript{4}}
\address{
  Nuance Communications, Inc., 
  $^2$Valencia, Spain,
  $^3$Torino, Italy,
  $^4$Burlington, MA, USA 
  }
\email {\{jesusandres.ferrer, dario.albesano, puming.zhan, paul.vozila\}@nuance.com}

\begin{document}

\maketitle
\begingroup\renewcommand\thefootnote{1}
\footnotetext{Both authors contributed equally to the work}
\begin{abstract}
    End-2-end (E2E) models have become increasingly popular in some ASR tasks  because of their  performance and advantages.
    These E2E models directly approximate the posterior distribution of tokens given the acoustic inputs. 
    Consequently, the E2E systems implicitly define a language model (LM) over the output tokens, which makes the exploitation of independently trained language models less straightforward than in  conventional ASR systems.
    This makes it difficult to dynamically adapt E2E ASR system to contextual profiles for better recognizing special words such as named entities. 
    In this work, we propose a contextual density ratio approach for both training a contextual aware E2E model and adapting the language model to named entities.
    We apply the aforementioned technique to an E2E ASR system, which transcribes doctor and patient conversations, for better adapting the E2E system to the names in the conversations.
    Our proposed technique achieves a relative improvement of up to 46.5\% on the names over an E2E baseline without degrading the overall recognition accuracy of the whole test set. 
    Moreover, it also surpasses a contextual shallow fusion baseline  by  22.1 \% relative. 
\end{abstract}

\section{Introduction}
 In recent years, End-to-end (E2E) systems~\cite{LAS,RNNT, NT}  have been greatly adopted for ASR because of their performance and simplicity. 
 In conventional ASR systems, several components, such as the pronunciation lexicon, the language models, and the acoustic model, are  optimized independently.
 In contrast, E2E systems   integrate all components of conventional ASR systems in a single neural network by directly approximating the posterior probability given the acoustic features.
 
 Despite the direct posterior approximation simplifying many aspects such as training, decoding and deployment, these advantages do not come without drawbacks. 
 Specifically, one of the disadvantages is the difficulty of exploiting external language models.
 Since the E2E models directly approximate the posterior token distribution, there is no straightforward way to integrate them, which 
  generates recognition problems for rare events among other problems.
  These tail probability events cannot be expected to be comprehensively observed in training.
The selection of the output units such as  word pieces~\cite{BPE} or characters instead of full words could theoretically mitigate these problems, but in practice this is not the case~\cite{ShallowContextual}.

 Several early attempts were  proposed to integrate external language models (LM) into E2E systems, ranging from shallow or deep fusion~\cite{ShallowDeepFusion, ShallowASR}, to cold fusion~\cite{coldFusion} among many others~\cite{componentFusion,ShallowFusionRNNT}.
 In the ASR domain, \cite{ShallowASR} shown that shallow fusion with same output units in both LM and E2E worked best to incorporate an external language model in some tasks.
 However, one of the  disadvantages of shallow fusion comes from the fact that most of the E2E systems~\cite{LAS,RNNT, LASFirst} arguably already incorporate   an internal LM~\cite{StatelessRNNT,LessIsMore, HAT}.
 Recently, a \emph{density ratio} approach~\cite{dr} was proposed to better integrate external LM by removing the internal LM contribution of E2E models while decoding.
 In this initial work~\cite{dr},  authors reported improvements over shallow fusion for domain adaptation tasks. Further works modified the architecture~\cite{HAT} or made some  assumptions~\cite{Internal} to better approximate the implicit LM of E2E systems in combination with density ratio.

 In many tasks, ASR systems are expected to recognize singletons or infrequent words such as contact list names or other named entities.
 These entities may not be seen during training, and even for those which are observed, the training data is clearly neither representative of the particular distribution nor of the relevant set associated with the application context for a given utterance.
 Consequently, even if the E2E model is using BPE~\cite{BPE} or graphemenes as output units, the system struggles to recognize them. Several approaches have been introduced in the literature to contextualize  E2E models to named entities such as song names or contact lists. In~\cite{CLAS} and~\cite{contextualRNNT}, an additional attention set input is proposed for both attention and recurrent neural transducer (RNN-T) models respectively. Shallow fusion approach~\cite{ShallowContextualBiasing} was applied surpassing state-of-the-art conventional models. More recently~\cite{ShallowContextual}, shallow fusion was used in combination with special tokens to delimit class-based entity names together with a mapping to transform  rare words to common words through pronunciation.

    In this paper, we propose \emph{contextual density ratio} for contextualizing E2E models so that the internal E2E LM is  dynamically adapted to a priori known named entities.
    The proposed technique builds upon both density ratio~\cite{dr} and class-based LM tags~\cite{ShallowContextual} to contextualize the E2E models. 
    During training, we introduce special tokens to enclose known named entities so that the E2E system learns to predict when a named entity is spoken 
    by the different statistical clues obtained from both the acoustics and the internal LM. 
    We also approximate an internal E2E language model by training an independent LM with the transcriptions on which the E2E system was trained.
    During decoding,  we apply density ratio within the named entities segments identified by the E2E model to dynamically adapt and contextualize the system's LM.
    We show that the proposed contextual density ratio (CDR) reduces names recognition errors over a  E2E baseline with and without contextual shallow fusion~\cite{ShallowContextualBiasing} by 46.5\% and 22.1 \% respectively.

    The paper is organized as follows. We first briefly review density ratio in section~\ref{sec:dr} to pave the way for the following section~\ref{sec:drt} where the proposed  contextual density ratio  is detailed.    
    In section~\ref{sec:results}, we apply the proposed technique to  a  doctor-to-patient conversation task for better recognizing both doctors and patient names. 
    Finally, in section~\ref{sec:conc}, we conclude with some reflections.

\section{Proposed approach}
\label{sec:proposed}
In this section we briefly summarize the \emph{contextual density ratio} approach. We first review the density ratio~\cite{dr}, and then we propose our extension  to the contextual biasing scenario.

\begin{figure*}[th]
   \addtolength{\tabcolsep}{-2pt}    
     \begin{tabular}{|c|ccccc|}
         \hline
         Model      &  mr\_ &{\tt\textless{}ne\textgreater}& moz & art\_ & {\tt\textless/ne\textgreater} \\ \hline
         E2E & 
                          $p(\text{mr\_} | \text{hello\_}, \vec{x})$ &
                          $p({{\tt\textless{}ne\textgreater}}|  \ldots,\text{mr\_}, \vec{x})$ &
                          $p(\text{moz} |  \ldots,  \text{mr\_}, {\tt\textless{}ne\textgreater}, \vec{x})$ &
                          $p(\text{art\_} |  \ldots,  \text{moz},\vec{x})$ &
                          $p( {\tt\textless/ne\textgreater}|  \ldots, \text{art\_},\vec{x})$ \\
         ID & $ 1 $& $ 1 $&
                          $p(\text{moz} |  \ldots, \text{mr\_},{{\tt\textless{}ne\textgreater}})$ &
                          $ p(\text{art\_} |  \ldots,  \text{moz})$ &
                          $ p( {\tt\textless/ne\textgreater}   | \ldots,  \text{art\_})$ \\
         NE &  $ 1 $& $ 1 $&
                          $ q(\text{moz} |  {\tt\textless{}ne\textgreater })$ &
                          $ q(\text{art\_} | {\tt\textless{}ne\textgreater }, \text{moz})$ &
                          $  q( {\tt\textless/ne\textgreater}   | {\tt\textless{}ne\textgreater },\text{moz,art\_}) $ \\
         \hline
     \end{tabular}
   \caption{An example of how the different components of contextual density ratio score a hypothesis. 
        The final hypothesis score is computed as  $\log(\ee) - \alpha\cdot\log(\id)  + \beta\cdot\log(\nent) $. 
        Note how the different language model components  track the context differently accordingly to their goal.
        The ID LM, is tracking full context from the beginning of the utterance in contrast with the NE LM which only tracks the LM 
        from the most recent  named entity start tag, {\tt <ne>}. 
    }
   \label{fig:dr}
 \end{figure*}

\subsection{Density ratio (DR)}
\label{sec:dr}
  Neural network E2E models such as~\cite{LAS, LASA}, directly approximate the probability of a target token sequence  $y=y_1^J$,  given acoustic features $\vec{x}=x_1^T$, as follows:
  \begin{equation}
      \label{eq:posterior}
       p(\vec{y}| \vec{x})\approx p_{\ee}(\vec{y}| \vec{x}) % = \prod_t p_{\ee}(y_t\mid \vec{x}, \vex{y}_{<t} )
  \end{equation}
  where  $I$ is the length of the input acoustic features along time and $J$ the length of a possible token sequence; and where $p$ denotes the actual probability of the token sequence given the input acoustic features and $p_{\ee}$ the approximated model distribution.

  In density ratio, we wish to adapt the posterior in-domain (ID) distribution $p(\vec{y}|\vec{x})$ to a new out-of-domain~(OOD) posterior, $q(\vec{y}|\vec{x})$. 
  Via standard noisy channel decomposition (Bayes’ law) and under assumption that acoustics of both domains are similar, 
  $
  p(\vec{x}|\vec{y}) \approx q(\vec{x}|\vec{y}) 
 $, we decompose the OOD posterior as:
  \begin{equation}
      \label{eq:noisyOOD}
       q(\vec{y}| \vec{x}) 
       = \frac{ q(\vec{x} | \vec{y}) q(\vec{y})}{q(\vec{x})}
       = \frac{ p(\vec{x} | \vec{y}) q(\vec{y})}{q(\vec{x})}
  \end{equation}
 Utilizing the same noisy channel decomposition for ID distribution $p(x|y)$ and rearranging terms, we obtain:
  \begin{equation}
      \label{eq:noisyID}
       p(\vec{x}| \vec{y}) = \frac{ p(\vec{y} | \vec{x}) p(\vec{x})}{p(\vec{y})}
  \end{equation}
  Plugging eq~\eqref{eq:noisyID} into eq.~\eqref{eq:noisyOOD} yields:
  \begin{equation}
      \label{eq:OOD_2}
       q(\vec{y}| \vec{x}) = \frac{ p(\vec{x} | \vec{y}) q(\vec{y})}{q(\vec{x})}
       = p(\vec{y} | \vec{x}) \frac{ q(\vec{y}) }{ p(\vec{y}) }  \frac{p(\vec{x})}{q(\vec{x})}
  \end{equation}
   Note that, eq.~\eqref{eq:OOD_2} has a marginal LM ratio $   q(\vec{y}) / p(\vec{y})  $ and an acoustics density ratio  $  p(\vec{x}) /q(\vec{x}) $, which does not modify token sequences scores.
  We obtain the final density ratio (DR) approach by approximating distributions with models:
  \begin{equation}
      \label{eq:DR}
       q(\vec{y}| \vec{x}) = p_{\ee}(\vec{y} | \vec{x}) \frac{ q_{\od}(\vec{y}) }{ p_{\id}(\vec{y}) }  \frac{p(\vec{x})}{q(\vec{x})}
    %   \approx p_{\ee}(\vec{x} | \vec{y}) \frac{ q_{\od}(\vec{y}) }{ p_{\id}(\vec{y}) } 
   \end{equation}
   which approximates the acoustic distribution with both an ID language and E2E  models, $p_{\id}$ and $p_{\ee}$;  and shifts the posterior with an OOD language model, $q_{\od}$, up to a constant ratio on the acoustics. 
   %
   % Removed because of disucssion with P. Vozila
   %Note that because $ p_{\id} $ corresponds to the marginal distribution obtained from the E2E model posterior, 
   % a good ID LM is one that better approximates the internal E2E LM.
   Note that $p_{\id}$ corresponds to the marginal distribution obtained from E2E posterior, $p_{\ee}$; in practice we take the E2E model training transcriptions as approximately sampling from this distribution.

   Finally, when performing the beam search we apply weights to balance out the importance of each score, yielding the search score:
  \begin{equation}
      \label{eq:OOD2}
       \operatorname{score}(\vec{y};\vec{x}) = 
       \log p_{\ee}(\vec{y} | \vec{x})  
       - \alpha\cdot\log p_{\id}(\vec{y})  
       + \beta\cdot\log q_{\od}(\vec{y}) 
   \end{equation}
 
   %, or token bonus~\cite{Shorowski2016-TBD,LASA}.
 
\subsection{Contextual density ratio (CDR)}
\label{sec:drt}

  Density ratio (DR)~\cite{dr} can bias the distribution for the full utterance, 
  but we would like to adapt the language model of E2E systems to be able to recognize named entities 
  such as contact list names, or client names.  These named entities may occur at any position inside a utterance and often only span a few words.

 In order to do so, we extend the token vocabulary with special words for both start and end of entities, {\tt <ne>} and  {\tt  </ne>}. 
 % %Below is duplicated
% Note that despite we focus on a single tag in this work,  we could add as many tags as  named entity types we want to cover. 
 %
 
 During training, our E2E model inserts those tags in between known named entities, for instance,  % {\tt call\_ mr\_ <ne> be e th o ve n\_ </ne>} or 
 {\tt the\_ next\_ patient\_ <ne> be e th o ve n\_ </ne> is\_ losing\_ his\_ hear ing\_ }. 
 The system learns to predict when those named entities will occur, 
 and during beam search or decoding, the system hypothesizes those special tokens as if they were standard tokens.
 Although we focus on a single tag for simplicity in this work, we could add as many specialized tags as desired, 

 In the previous section, we reviewed the DR approach for full utterance biasing. However, in practice, each token distribution on eq.~\eqref{eq:DR} is approximated autoregressively as follows:
  \begin{equation}
      \label{eq:posterior_t}
       q(\vec{y}| \vec{x}) = \frac{p(\vec{x})}{q(\vec{x})} \prod_t  p_{\ee}(y_t|\vec{x}, \vec{y}_{<t} ) \frac{q_{\od}(y_t|\vec{y}_{<t})}{p_{\id}(y_t|\vec{y}_{<t})}
  \end{equation}
Because of the new tokens added to the E2E vocabulary,  we know the positions where the named entities will occur. 
For simplicity, we assume a single named entity occurrence at position $(t_b,t_e)$.
Then, the token posterior probability of an utterance with a named entity is decomposed by:
  \begin{equation}
     \label{eq:CDR}
      \begin{split}
          q_{\text{ne}}(\vec{y}| \vec{x}) =& \left(\prod_{t<t_b} p_{\ee}(y_t| \vec{x}, \vec{y}_{<t} )\right) \\
          &\cdot p_{\ee}(\text{\tt<ne>}| \vec{x}, \vec{y}_{<t_b} ) \\
          & \prod_{t=t_b+1}^{t_e} p_{\ee}(y_t | \vec{y}_{< t},\vec{x})  \frac{ q_{\nent}(y_t| \vec{y}_{t_b}^{t-1}) }{ p_{\id}(y_t|\vec{y}_{<t} ) }
           \frac{p(\vec{x})}{q(\vec{x})}\\
          %& p_{\ee}(\text{{\tt</ne>}}| \vec{x}, \vec{y}_{<e_2} )
          & \prod_{t>t_e} p_{\ee}(y_t| \vec{x}, \vec{y}_{<t} )   
      \end{split}
  \end{equation}
  where recall $y_{t_e}=\text{\tt</ne>}$. 
  Note that the in-domain (ID) and named entity (NE) LMs condition on different context.
  It is worth mentioning that the acoustic density ratio, $  p(\vec{x}) /q(\vec{x}) $, may introduce a bias when comparing hypotheses with different number of named entities.
  However, assuming acoustics are similar, this ratio should approach 1 and we verified in the next section that this is not a problem in practice. 

  During beam search or decoding, each token is scored with  $ \operatorname{score}(y_t) :=  \operatorname{score}(y_t; \vec{y}_{<t}, \vec{x}) $ computed as:
  \begin{equation}
      \label{eq:OOD2}
       \operatorname{score}(y_t) \!\!= \!\!
       \begin{cases}
           \log p_{\ee}(y_t |\ldots) + 
           \log \frac{ q_{\nent}(y_t | \vec{y}_{t_b}^{t-1})^\beta} { p_{\id}(y_t | \vec{y}_{<t})^\alpha}, &\!\!\!\! \operatorname{ne}(\vec{y}_{<t}) \\
           \log p_{\ee}(y_t |\ldots),  & \!\!\!\! \text{o.w.}
       \end{cases}
   \end{equation}
  where we  omit $\vec{y}_{<t},\vec{x}$ from $p_{\ee}(y_t|\ldots)$. The notation $  \operatorname{ne}(\vec{y}_{<t}) $  indicates whether there is an open named entity tag in the previous  tokens that has not yet been  closed and in such a case, $t_b := t_b(\vec{y}_{<t})$ is the position of the start tag token, {\tt<ne>}. 
  In practice,  LM distributions are weighted by  $\alpha$ and $\beta$ to smooth them.
  
  Figure~\ref{fig:dr} depicts a small segment of an utterance  with a named entity. 
  The E2E model scores each token independently of whether it is within a named entity tag, in contrast to both in-domain (ID) and LM named-entity (NE) LM.
  Since the ID LM  is subtracted to counteract the internal E2E LM, it is tracking the context of the full utterance, whereas 
  the NE LM, only tracks the context within the contextual named entity tags.

\section{Experiments}
\label{sec:results}

\subsection{Data set}

We apply the proposed contextual density ratio to a conversational speech transcription task (doctor-patient conversations).
All speech data are field data, from which we computed 80 filter bank features from the input audio.
For this study, the train and test set consisted of 1.5K hours and 26.8 hours respectively from several medical domains.
%In addition, the test set is composed of 21K sentences and 296K words from several medical domains.

We focused on person names as named entities and automatically marked them in both training and testing (references) by intersecting the transcriptions with a list of 173K English names. 
During training we enclosed  occurrences of such names between tags ({\tt\textless{}ne\textgreater} and {\tt\textless{}/ne\textgreater}) and during testing we extracted for each conversation both doctor and patient names from conversation utterances. 
In total 5\% of the utterances in the test set have names.
%There are a total of 1\,260 names (622 unique names) in the test set distributed across 1063 utterances and 256 encounters; 4.95\% of the utterances have names. 

% While mechanism above to extract names may sound having Oracle knowledge, 
% in production we have access to such information and as we will 
% see in the experimental section below,  we also verified  our results in actual production 
% training data by training model in a larger 
% dataset and  making use encounter names.

\subsection{Models}
  The E2E model is a fairly standard E2E attention model similar to LAS~\cite{LAS}.
  Specifically, our model follows the same architecture of the speaker independent system from LASA~\cite{LASA}.
  Our model has a set of 3 convolutional neural network (CNN) layers which are configured to produce a 768-dimensional embedding for each input frame. The 3 CNN layers have a kernel size of 3 and do not perform any time reduction. Subsequently, there is a pyramid LAS-like  encoder composed of 6 bidirectional Long Short-Term Memory (bLSTM)~\cite{bLSTM} layers with 768 hidden units per layer 
  and direction. 
  The encoder performs input decimation by a factor of 2 after every other layer.
  The total time reduction is 8x with respect to the filter banks inputs. 
  The decoder component uses 2 unidirectional LSTM~\cite{LSTM} layers of 1536 units per layer. We use Bahdanau attention in an architecture similar to Luong~\cite{Luong,LASA} on the attention component, with a single head with 1536 nodes. 
  We apply dropout~\cite{Dropout} to regularize the layers and SpecAugment~\cite{SpecAugment} 
  to both regularize the model and as data augmentation. We use a single time band of length 40 frames and a 
  frequency band of up to 20 filters. The final softmax layer predicts the posterior probability of $2.5$K word-pieces~\cite{BPE} computed on the training set. The model accounts for a total of $143.7$M parameters.
  
  In order to approximate the internal E2E language model for the contextual density ratio technique, we use the exact transcriptions of the utterances on which the E2E model was trained,  together with the very same $2.5$K word-pieces lexicon.
  We trained the language model with truncated backpropagation through time (BPTT)~\cite{BPTT} to better approximate the internal E2E language model.
  We settled for a language model with 512 embedding projection followed by 2 unidirectional LSTM layers of 1536 each and a final down projection back to 512 logits followed by the softmax layer of $2.5$K outputs. The internal LM accounts for a total of  31.99M parameters. We regularized the model with small weight-decay and dropout. 
  %We attempted to use stronger LM with larger parametrization and also smaller language models but all performed worse,
  %which enforce the theoretical insight that the in-domain LM, $p_{\id}(y_t | \vec{y}_{<t})$, needs to correctly approximate 
  %the internal E2E LM. 
  For decoding, both the $\alpha$ and $\beta$ parameters were tuned on a different domain adaptation density ratio setup and both equal to $0.1$. 
   This ensures that search parameters are not over-trained to the specifics of the contextual density ratio task or the test set. In addition, because of the attention models, we used a coverage score~\cite{Shorowski2016-TBD} of $0.3$ optimized in conjunction with $\alpha$ and $\beta$.
  
  For the contextual LMs, we initialized them with the same in-domain LSTM based language model and trained them until  convergence  on the specific list of names extracted per each conversation.

\subsection{Results}

   In order to measure the different systems, we report the word error rate (WER), as well as  the WER within the name tags (WERT)\footnote{We aligned hypotheses and references with tags and computed the errors aligned with the references tags.}.
   This allow us to evaluate both overall performance of the system as well as the specific  improvements on name recognition.
  
  First, we trained both an E2E baseline model without tags and an E2E model with the name tags ({\tt\textless{}ne\textgreater} and {\tt\textless{}/ne\textgreater}). Second line of Table~\ref{tab:wer} shows 
  that adding the tags slightly hurts the model performance, compared to the first line, but 
   the difference is small and within training repetition variance (+/- 0.1). 
  Then, we added contextual biasing by either shallow fusion~\cite{ShallowContextual} or contextual density ratio in the realistic  scenario, in which  we build a context biasing language model per  conversation. 
  As reported in the last 2 lines of Table~\ref{tab:wer}, in both cases, we recover  part of the small drop in accuracy while significantly improving name performance.
  Adding a contextual biasing shallow fusion component reduces name errors by 31.3 \% relative.
  Contextual density ratio biasing  reduces name errors by 46.5 \% relative with respect to the E2E baseline and 22.1 \% compared to shallow fusion. %
   Note that if all names were wrongly recognized by substitutions, this will account for only 0.43\%  of the whole test set WER.
  
  \begin{table}[ht]
     \caption{
    Word Error Rate (WER) and Word Error Rate within the named entities Tags (WERT) for several systems. 
    Both contextual systems based on either SF or DR, extend the E2E model with contextual tags and are based on full conversations.
    }
    \centering
    \begin{tabular}{|c|cc|}\hline
        \bf System                  &     \bf WER & \bf WERT            \\\hline
        E2E without contextual tags            &     {\bf 14.94 }    &    n.a.             \\\hline
        E2E with contextual tags               &     15.10    &    44.1             \\
        \phantom{E2E}+ contextual SF \phantom{(S DR)}  &     15.04    &    30.3             \\
        \phantom{E2E}+ contextual DR    (CDR)       &     {\bf 15.00}    &    {\bf 23.6}             \\\hline
    \end{tabular}
 
    \label{tab:wer}
\end{table}

Table~\ref{tab:encounter}, compares the precision and the recall on the contextual tags themselves for the 3 systems with tags from Table~\ref{tab:wer}.
 Both contextual density ratio and contextual shallow fusion improve the tagging precision and recall over the baseline, with contextual density ratio being  better than contextual shallow fusion. 
 The improvements in terms of precision and recall are small compared to the WERT improvements from Table~\ref{tab:wer}, and consequently the name WERT improvement cannot be attributed as a side effect of  better tagging. 
 We hypothesize instead that the a priori known entity values for the conversation, allow the contextually biased techniques to better detect the span for such entities, in particular when linguistic contextual cues are insufficient.
  
\begin{table}[t]
    \centering
    \caption{Precision and recall computed on the contextual tags themselves for the different contextual systems on table~\ref{tab:wer}. }
    \begin{tabular}{|l|cc|}\hline
        \bf System                                     & \bf Precision  & \bf Recall \\\hline
        E2E with contextual tags                &  78.5    & 79.4 \\ 
%                  +Tags                          &  {\color{red}44.1}  &{\color{red}    78.5}        & {\color{red} 79.4} \\\hline
  \phantom{E2E}+ contextual SF \phantom{(S DR)} &   80.7      &  82.6\\
        \phantom{E2E}+ contextual DR    (CDR)       &   {\bf 82.0}       & {\bf 83.5 }\\\hline
    \end{tabular}
    \label{tab:encounter}
\end{table}

\subsection{Oracle and noisy names list experiments}
We studied the proposed contextual DR (CDR) in several scenarios. 
Initially, we evaluated  the oracle best case where we  know whether a given utterance has a name and 
the name itself in such a case. For those utterances with names, a contextual name LM is trained per utterance.

This oracle setup bounds the potential improvements with respect to the more realistic experiments from the previous section, and measures the effect of the acoustic ratio in Eq.~\eqref{eq:CDR} because no utterance without names is decoded with the biasing component.
CDR applied to full conversations,  is  close to the oracle case as shown in first row of Table~\ref{tab:oracle}.
On the other extreme, a single contextual name LM was trained on all names in the full test set as if the conversation level information was lost. The second row of Table~\ref{tab:oracle} shows that using a single LM has a larger negative effect than using conversation-specific LMs, yet it is still better than the E2E baseline
%in Table~\ref{tab:wer}.

\begin{table}[t]
  \caption{ 
    WERT precision and recall for a single LM per utterance (oracle case) and a single LM for the full test set dropping conversation level information.
    }
    \centering
    \begin{tabular}{|l|c|c@{\hspace{1.0\tabcolsep}}c@{\hspace{0.5\tabcolsep}}|}\hline
       \bf System                                   & \bf WERT  &  \bf Precision  &\bf Recall \\\hline
        1 LM per utterance (Oracle)                 &  20.1  & 82.6       & 85.1  \\
        %1 LM with 1260 names
        1 LM for full test set 
        &  37.3   & 78.5      & 79.7 \\\hline
    \end{tabular}
  
    \label{tab:oracle}
\end{table}

Following the worst case experiment from Table~\ref{tab:oracle}, 
we studied the effect of extending the entity (or name)  lists  per each conversation,  by adding an increasing number of randomly sampled distracting names (from 1 to 256) unrelated to the conversation.
As shown in Figure~\ref{fig:confusion}, CDR is relatively robust to noisy names,
only starting to degrade to 26.0\% WERT at 256 distracting names. 
Specifically, for 16 distracting names we obtain a WERT of 23.9 which is very close to 23.6 of the conversation level LMs in Table~\ref{tab:wer}. 
%Note there is small statistical variation between runs because of the contextual LM trainings.

Finally,  the entity list per conversation, which is used to train the CDR contextual LMs,  was extended with 16 adversarial names so as to further investigate the robustness of CDR.
These names are randomly selected from a pool of 170K names (and their combinations) according to their character lexicographical Levensthein distance, in \{1,2,4\}, from any of the actual conversation names.
Note that for Levensthein distance 4, we can drop part of the name (e.g. in case of John Smith the resulting adversarial name can be Smith).
Table~\ref{tab:adversarial} assesses how adversarial examples degrade the performance of the CDR biasing; it is still relatively robust and much better than that of the baseline system. Note that the performance drop for edit distance 4 may seem unexpected, but it is in part a side effect of random sampling being able to drop full short names or surnames via deletions, in contrast to 2 or 1, where names are rarely dropped if at all. 
We intentionally allowed for full name deletions so as to make the list insidiously difficult.

\begin{table}[t]
    \caption{ WERT for 16 names with different Levensthein distances, applied to per-conversation CDR case. }
    \centering
    \begin{tabular}{|l|cccc|}\hline
       \bf Levensthein distance & \bf random & \bf 4    &  \bf 2   & \bf 1     \\\hline
       WERT          & 23.9   & 28.2 & 30.5 & 28.3  \\\hline
    \end{tabular}
    \label{tab:adversarial}
\end{table}

\begin{figure}[th]
   \centering
   \includegraphics[width=\linewidth]{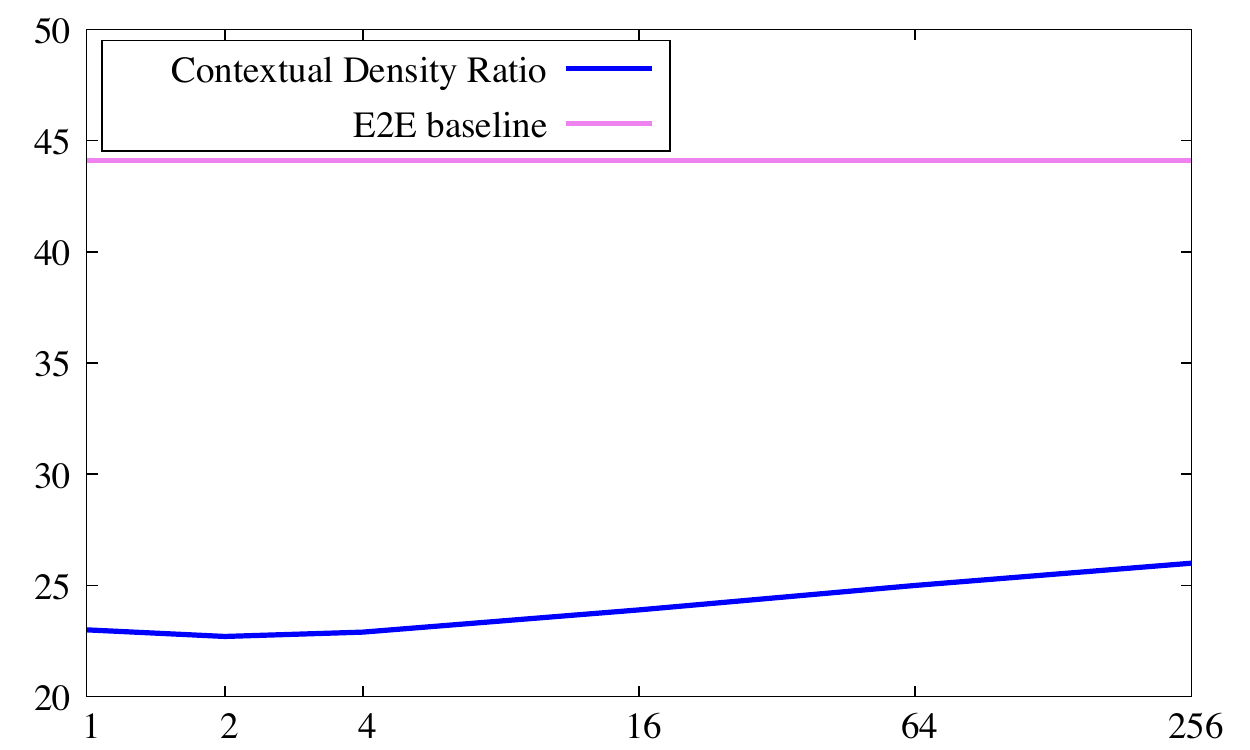}
   \caption{WERT (y-axis) at increasing number of per-conversation randomly sampled distracting names (log-scale).}
   \label{fig:confusion}
 \end{figure}

\section{Conclusions}
\label{sec:conc}
 In this paper, a contextual density ratio for contextual language model biasing was proposed.
 This technique was applied to the task of name recognition in doctor-patient conversations.
 The proposed approach improves name recognition up to 46.5\%  with respect to a standard E2E system or
 22.1\% relative with respect to contextual shallow fusion.  Moreover, the  technique does 
 not degrade the system performance on utterances without names significantly and has no major side effects.
 
 The behaviour of the proposed technique was studied by adding random distracting and adversarial names to the biasing name list. Contextual density ratio is robust to noise, being more sensitive to  similar names.
 As  future work, we want to improve the  technique  in those adversarial conditions, for instance by also taking into account the phonetics.

%\section{Acknowledgements}

\bibliographystyle{IEEEtran}

\bibliography{mybib}

\end{document}